\documentclass[11pt]{article}
\usepackage[margin=2cm]{geometry}
\usepackage{comment}
\usepackage{amsmath,amssymb,extarrows,mathtools,graphicx,subfigure,setspace}
\usepackage{cite}
\usepackage{slashed}
\usepackage{color}
\usepackage{amsmath,amsthm,amsfonts}

\usepackage{fullpage}
\makeatother

\numberwithin{equation}{section}

\newcommand{\be}{\begin{equation}}
\newcommand{\bea}{\begin{eqnarray}}
\newcommand{\eea}{\end{eqnarray}}
\newcommand{\ba}{\begin{array}}
\newcommand{\ea}{\end{array}}
\newcommand{\ee}{\end{equation}}

\def\ba{\begin{array}}
\def\ea{\end{array}}
\def\be{\begin{equation}}
\def\ee{\end{equation}}

\def\-1{^{-1}}

\input{xy}
\xyoption{all}
\xyoption{matrix}

\theoremstyle{plain}

\date{}

\begin{document}
\onehalfspacing
\noindent
\begin{titlepage}
\hfill
\vspace*{20mm}
\vspace{-8mm}
\begin{center}
{\Large {\bf BTZ black hole from Poisson-Lie T-dualizable sigma models\\ with spectators}\\
}

\vspace*{15mm} \vspace*{1mm} A. Eghbali~\hspace{-1mm}{\footnote{Corresponding author: a.eghbali@azaruniv.edu}}\hspace{-1mm},
L. Mehran-nia  and A. Rezaei-Aghdam \hspace{-2mm}{ \footnote{ rezaei-a@azaruniv.edu}}
\\

{\it  Department of Physics, Faculty of Basic Sciences,\\
 Azarbaijan Shahid Madani University, 53714-161, Tabriz, Iran \\ }

\vspace*{.4cm}

\vspace*{.4cm}

\end{center}
\begin{abstract}
The non-Abelian T-dualization of the BTZ black hole is discussed in detail by using
the Poisson-Lie T-duality in the presence of spectators.
We explicitly construct a dual pair of  sigma models related by Poisson-Lie symmetry. The original model is built
on a $2+1$-dimensional manifold ${\cal M} \approx O \times \bf G$, where $\bf G$ as a  two-dimensional real non-Abelian Lie group
acts freely on ${\cal M}$, while $O$ is the orbit of $\bf G$ in ${\cal M}$.
The findings of our study show that the original model indeed is canonically equivalent to
the $SL(2,\mathbb{R})$ Wess-Zumino-Witten (WZW) model for a given value of the background parameters.
Moreover, by a convenient coordinate transformation we show that this model
describes a string propagating in a spacetime with the BTZ black hole metric
in such a way that a new family of the solutions to low energy string theory with the BTZ black hole
vacuum metric, constant dilaton field and a new torsion potential is found.
The dual model is built on a $2+1$-dimensional target
manifold $\tilde {\cal M}$ with two-dimensional real Abelian Lie group ${\tilde {\bf G}}$ acting freely on it.
We further show that the dual model yields a three-dimensional charged black string for which
the mass $M$ and axion charge $Q$
per unit length are calculated. After that,  the structure and  asymptotic nature of the dual
space-time  including the horizon and singularity are determined.
\end{abstract}


\end{titlepage}


\section{\large Introduction}

2+1-dimensional black hole solution with a negative cosmological constant, mass, angular momentum
and charge was, first, found by Banados, Teitelboim and  Zanelli (BTZ) \cite{Banados}.
The study of 2+1-dimensional solutions have received a lot of attention, since
the near horizon geometry of these solutions serves as a worthwhile model
to investigate some conceptual questions of $AdS/CFT$ correspondence \cite{Carlip}, \cite{Witten1}.
The BTZ black hole is asymptotically anti-de Sitter rather than asymptotically flat, and has no curvature singularity
at the origin. A slight modification of this black hole solution
yields an exact solution to string theory \cite{Horowitz2}.

One of the most interesting properties of string theories, or more specially, two-dimensional non-linear sigma models
is that a certain class of these models admits the action of duality transformations which
change the different spacetime geometries but leave unchanged the physics of such theories at the
classical level.
Witten had shown \cite{Witten2} that two-dimensional black hole could be obtained by
gauging a one-dimensional subgroup of
$SL(2,\mathbb{R})$. Then, it was  shown \cite{Horowitz1} that a simple
extension of Witten's construction yields a three-dimensional charged black string such that this
solution is  characterized by three parameters the mass, axion charge
per unit length, and a constant $k$ \cite{Horowitz1}.
After all,  it was shown \cite{Horowitz2} that the BTZ black hole is, under the standard Abelian T-duality, equivalent to
the charged black string solution discussed in Ref. \cite{Horowitz1}.
In the present paper, we obtain this result by making use of the approach of the Poisson-Lie T-duality
with spectators \cite{Klimcik1},  \cite{Tyurin}, \cite{Sfetsos1}.
The Poisson-Lie T-duality is a generalization of Abelian \cite{Buscher1} and
traditional non-Abelian dualities \cite{nonabelian} that proposed by  Klimcik and Severa \cite{Klimcik1}.
It deals with sigma models based on two Lie groups which form a Drinfeld
double and the duality transformation exchanges their roles.
This duality is a canonical transformation and two sigma models related by Poisson-Lie
duality are equivalent at the classical level \cite{Sfetsos2}.
A generalization of the Poisson-Lie T-duality transformations
from manifolds to supermanifolds has been also carried out in Ref. \cite{ER1} (see, also, \cite{ER5}).
One of the most interesting applications of the Poisson-Lie T-duality transformations is to
the WZW models \cite{Alekseev1}, \cite{eghbali11}. Up to now, only few examples of the Poisson-Lie  symmetric sigma models  have been treated at the
quantum level \cite{Alekseev1}, \cite{Lledo}.
Furthermore, the Poisson-Lie symmetry in the WZW models based on the Lie  supergroups have recently studied
in Refs. \cite{ER7} and \cite{ER8}.
We also refer the reader to the literatures \cite{Sfetsos3}.
In \cite{Alekseev1} it has been shown that the duality relates the $SL(2 , \mathbb{R})$ WZW model to a constrained
sigma model defined on the $SL(2,\mathbb{R})$ group space.
Here in this paper we obtain a new non-Abelian T-dual background for the $SL(2 , \mathbb{R})$ WZW model so that it
is constructed out on a $2+1$-dimensional manifold.
Moreover, by using the Poisson-Lie T-duality with spectators
we find a new family of the solutions to low energy string theory with the BTZ black hole
vacuum metric, constant dilaton field and a new torsion potential.
The findings of our study show that the dual model yields a three-dimensional charged black string. This
solution is stationary and is characterized by three parameters: the mass $M$ and axion charge $Q$
per unit length, and a radius $l$ related to the derivative of the asymptotic value of the
dilaton field. In this way, we show that the non-Abelian T-duality
transformation (here as the Poisson-Lie T-duality on a semi-Abelian double) relates a solution with no horizon
and no curvature singularity (the BTZ vacuum solution) to a solution with a single horizon and a curvature singularity
(the charged black string).

To set up conventions and notations, as well as to make the paper self-contained, we review the
Poisson-Lie T-dual sigma models construction in the presence of spectators in Section 2.
In Section 3 we discuss in detail the Abelian T-dualization of
the BTZ black hole solutions by using the approach of the Poisson-Lie T-duality with spectators.
Our main result is derived in Section 4:
we explicitly construct a dual pair of  sigma models related by Poisson-Lie symmetry on $2+1$-dimensional manifolds ${\cal M}$ and
$\tilde {\cal M}$ such that the group parts of ${\cal M}$ and $\tilde {\cal M}$ are two-dimensional real non-Abelian and Abelian Lie groups, respectively.
In subsection 4.1, by a convenient coordinate transformation we show that
the original model describes a string propagating in a spacetime with BTZ black hole
vacuum metric. Moreover, the canonical equivalence of the original model  to
the $SL(2,\mathbb{R})$ WZW model is discussed in subsection 4.1.
In subsection 4.2, first the dual model construction is given. Then
it is shown that the dual model yields a three-dimensional charged black string
for which the mass $M$ and axion charge $Q$ per unit length are calculated.
Investigation of the structure and  asymptotic nature of the dual space-time
including the horizon and singularity are  discussed  at the end of Section 4.
Some concluding remarks are given in the last Section.

\vspace{-2mm}
\section{\large A review of  Poisson-Lie T-duality with spectators}
We start this section by recalling the main features of the Poisson-Lie T-duality
and introducing redefinitions to make the duality transformations more
symmetrical. According to \cite{Klimcik1}, the Poisson-Lie duality is based on
the concepts of the Drinfeld double which is simply a Lie group $\bf D$, such that the Lie algebra ${\cal D}$ of this Lie group as a vector space
can be decomposed into a pair of maximally isotropic subalgebras ${\cal G}$ and ${\tilde {\cal G}}$ of the Lie groups $\bf G$ and ${\tilde {\bf G}}$, respectively.
We take the sets $T_a$, $a=1,...,dim (\bf G)$, and ${\tilde T}^a$  as the
basis of the Lie algebras ${\cal G}$ and ${\tilde {\cal G}}$, respectively. They satisfy the commutation relations
\begin{eqnarray}\nonumber
~~~[T_a , T_b] = {f^c}_{ab} ~T_c,~~~~~~~~[{\tilde T}^a , {\tilde T}^b] ={{\tilde f}^{ab}}_{\; \; \: c} ~{\tilde T}^c,
\end{eqnarray}
\vspace{-11mm}
\begin{eqnarray}\label{a.1}
[T_a , {\tilde T}^b] = {{\tilde f}^{bc}}_{\; \; \; \:a} {T}_c + {f^b}_{ca} ~{\tilde T}^c.~~~~~~~~~~~~~~~
\end{eqnarray}
In addition to \eqref{a.1}, there is an inner product $<. , .>$ on ${\cal D}$  with
the various generators obeying
\begin{eqnarray}\label{a.2}
<T_a , {\tilde T}^b> ~=~ {\delta _a}^{~b},~~~~~    <T_a , T_b> ~= ~<{\tilde T}^a, {\tilde T}^b> ~ =~ 0.
\end{eqnarray}
In what follows we shall investigate the Poisson-Lie T-duality transformations with spectators \cite{Klimcik1},
\cite{Sfetsos1} of non-linear sigma models given by
the action
\begin{eqnarray}\label{a.3}
S ~=~ \frac{1}{2} \int_{\Sigma} d\sigma^{+}  d\sigma^{-} ~ \big[{G}_{_{\Upsilon\Lambda}}(\Phi) +
{ B}_{_{\Upsilon\Lambda}}(\Phi)\big] ~ \partial_{+} \Phi^{^{\Upsilon}} \partial_{-} \Phi^{^{\Lambda}},
\end{eqnarray}
where $\sigma^{\pm} = \tau \pm \sigma$ are the standard light-cone variables on the worldsheet ${\Sigma}$.
${G}_{_{\Upsilon\Lambda}}$ and ${B}_{_{\Upsilon\Lambda}}$  are components of the metric and antisymmetric tensor field
${B}$ on a manifold ${\cal M}$. The functions $\Phi^{^{\Upsilon}}:{\Sigma} \longrightarrow \mathbb{R}$, ${\Upsilon} =1,...,dim({\cal M})$
are obtained by the composition $\Phi^{^{\Upsilon}}={X^{^{\Upsilon}}} \circ \Phi$ of a map $\Phi:{\Sigma} \longrightarrow {\cal M}$ and components of a coordinate map ${X}$ on a chart of ${\cal M}$.

Let us now consider a $d$-dimensional manifold ${\cal M}$ and some coordinates $\Phi^{^{\Upsilon}} = \{x^{\mu} , y^\alpha\}$ on it,
where $x^\mu,~\mu = 1,\cdots,dim (\bf G)$
are the coordinates of Lie group $\bf G$ that act freely from right on ${\cal M}$. $y^\alpha$ with $\alpha = 1,\cdots,d-dim (\bf G)$ are
the  coordinates labeling the orbit $O$ of $\bf G$ in the target space ${\cal M}$.
We note that  the coordinates $y^\alpha$ do not participate in the Poisson-Lie T-duality transformations
and are therefore called spectators \cite{Sfetsos1}.
We also introduce the components of the right invariant one-forms $(\partial_{_{\pm}} g g^{-1})^a = {R_{_{\pm}}^a} =
\partial_{_{\pm}} x^\mu R_{\mu}^{~a}$
and for notational convenience we will also use ${R_{_\pm}^{\alpha}} = \partial_{_{\pm}} y^{\alpha}$. Then the
original sigma model on the manifold ${\cal M} \approx O \times \bf G$ is defined by the following action
\begin{eqnarray}\label{a.27}
S &=&\frac{1}{2} \int d\sigma^{+}  d\sigma^{-}  ~{\mathbb{F}_{_{AB}}^{+}}(g,y^{\alpha})~{R_+^A}\;{R_-^B},\nonumber\\
&=&\frac{1}{2} \int d\sigma^{+}  d\sigma^{-}\Big[{\mathbb{E}_{_{ab}}^{+}}(g,y^{\alpha})~
{R_+^a}\;{R_-^b}+\Phi^{+^{(1)}}_{a \beta}(g,y^{\alpha}) {R_+^a}\partial_{-} y^{\beta}\nonumber\\
&&~~~~~~~~~~~~~~~~~~~~~~~~~~~~~~~~~+
\Phi^{+^{(2)}}_{\alpha b}(g,y^{\alpha}) \partial_{+} y^{\alpha} {R_-^b}+\Phi_{_{\alpha \beta}}(y^{\alpha})
\partial_{+} y^{\alpha} \partial_{-} y^{\beta}\Big],
\end{eqnarray}
where the index $A=\{a, \alpha\}$. As it is seen the couplings ${\mathbb{E}_{_{ab}}^{+}}$, $\Phi^{+^{(1)}}_{a \beta}$, $\Phi^{+^{(2)}}_{\alpha b}$
and  $\Phi_{_{\alpha \beta}}$ may depend on all variables $x^\mu$ and $y^\alpha$. The background matrix ${\mathbb{F}_{_{AB}}^{+}}(g,y^{\alpha})$
is defined as
\begin{eqnarray}\label{a.11}
\mathbb{F}^{+}_{_{AB}}(g,y^{\alpha})\;=\;A_{_{A}}^{^{~C}}(g)~E^{+}_{_{CD}}(g,y^{\alpha})~A_{_{B}}^{^{~D}}(g),
\end{eqnarray}
where \cite{Klimcik1}
\begin{eqnarray}\label{a.9}
E^{+}_{AB}(g,y^{\alpha})\;=\;{\big(A(g) + E^{+}(e,y^{\alpha}) B(g)\big)^{-1}}_{_{\hspace{-3mm}A}}^{^{~C}}~E^{+}_{_{CD}}(e,y^{\alpha})~
(A^{-1})_{_{B}}^{^{~D}}(g),
\end{eqnarray}
with $e$ being the unit element of $\bf G$. Here, the matrix $E^{+}(e,y^{\alpha})$ may be function of the spectator variables $y^\alpha$ only and
is defined in terms of the new couplings $E^{+}_{0\;ab}$, $F^{+^{(1)}}_{a \beta}$, $ F^{+^{(2)}}_{\alpha b}$ and
$F_{\alpha \beta}$ so that it is read
\begin{eqnarray}\label{a.6}
E^{+}_{_{AB}}(e,y^{\alpha})\;=\;\left( \begin{array}{cc}
                     E^{+}_{0\;ab}\tiny(e,y^{\alpha}) & F^{+^{(1)}}_{a \beta}(e,y^{\alpha})\\

                     F^{+^{(2)}}_{\alpha b}(e,y^{\alpha}) & F_{\alpha \beta}(y^{\alpha})
                      \end{array} \right).
\end{eqnarray}
In addition, the matrices $A(g)$ and $B(g)$ appearing in relations \eqref{a.11} and \eqref{a.9} are given by\footnote{Here Id means the identity matrix.}
\begin{eqnarray}\label{a.10}
A(g)\;=\;\left( \begin{array}{cc}
                     a(g) & 0\\

                     0 & Id
                      \end{array} \right),~~~~~~~~~
                      B(g)\;=\;\left( \begin{array}{cc}
                     b(g) & 0\\

                     0 & 0
                      \end{array} \right),
\end{eqnarray}
where the submatrices $a(g)$ and $b(g)$ associated with the Lie group $\bf G$ are constructed using
\begin{eqnarray}\label{a.7}
g^{-1} T_{{_a}}~ g ~=~ a_{_{a}}^{^{~c}}(g) ~ T_{{_c}},~~~~~~~~g^{-1} {\tilde T}^{{^a}} g ~=~
b^{^{ac}}(g)~ T_{{_c}}+(a^{-1})_{_{c}}^{^{~a}}(g)~{\tilde T}^{{^c}},
\end{eqnarray}
and for later use we also need to consider the following definition
\begin{eqnarray}
\Pi^{^{ab}}(g) ~=~ b^{^{ac}}(g)~ (a^{-1})_{_{c}}^{^{~b}}(g).\label{a.8}
\end{eqnarray}
Thus, using the relations \eqref{a.11}-\eqref{a.10} together with \eqref{a.8} one can obtain the backgrounds appearing
in the action \eqref{a.27}. They are then given in matrix notation by
\begin{eqnarray}
{\mathbb{E}}^{+}\tiny(g,y^{\alpha})&=&\Big(E^{+^{-1}}_{0}\tiny(e,y^{\alpha})+ \Pi(g)\Big)^{-1},\label{a.13}\\
\Phi^{+^{(1)}}(g,y^{\alpha})&=& {\mathbb{E}}^{+}\tiny(g,y^{\alpha})~{E^{+}_{0}}^{-1}\tiny(e,y^{\alpha})~F^{+^{(1)}}(e , y^{\alpha}),\label{a.14}\\
\Phi^{+^{(2)}}(g,y^{\alpha})&=& F^{+^{(2)}}(e , y^{\alpha}) ~{E^{+}_{0}}^{-1}\tiny(e,y^{\alpha})~{\mathbb{E}}^{+}\tiny(g,y^{\alpha}),\label{a.15}\\
\Phi(y^{\alpha})&=& F(y^{\alpha})-F^{+^{(2)}}(e , y^{\alpha})~\Pi(g)~{\mathbb{E}}^{+}\tiny(g,y^{\alpha})~{E^{+}_{0}}^{-1}\tiny(e,y^{\alpha})~F^{+^{(1)}}(e , y^{\alpha}).\label{a.16}
\end{eqnarray}
As we shall see below, one can construct another sigma model (denoted as usual
with tilded symbols) which is said to be dual to \eqref{a.27}
in the sense of the Poisson-Lie T-duality if the Lie algebras $\cal G$ and ${\cal \tilde G}$ form a pair of maximally isotropic
subalgebras of the Lie algebra $\cal D$ \cite{Klimcik1}.
Similarly to \eqref{a.6} one can define the matrix ${{\tilde E}^{+}}({\tilde e} , y^{\alpha})$ and then obtain the
relation between both the matrices $E^{+}(e,y^{\alpha})$ and ${{\tilde E}^{+}}({\tilde e} , y^{\alpha})$ as follows \cite{Klimcik1}
\begin{eqnarray}\label{a.29}
{{\tilde E}^{+}}({\tilde e} , y^{\alpha}) ~=~\big({\cal A}+{{E}^{+}}({e} , y^{\alpha})~{\cal B}\big)^{-1}~
\big({\cal B} + {{E}^{+}}({e} , y^{\alpha})~{\cal A}\big),
\end{eqnarray}
in which
\begin{eqnarray}\label{a.30}
{\cal A}\;=\;\left( \begin{array}{cc}
                     0 & 0\\

                     0 & Id
                      \end{array} \right),~~~~~~~~~
                    {\cal B}\;=\;\left( \begin{array}{cc}
                     Id & 0\\

                     0 & 0
                      \end{array} \right).
\end{eqnarray}
Now, the dual sigma model on the manifold ${\tilde {\cal M}} \approx O \times {\tilde {\bf G}}$ is,
in the coordinate base  $\{{\tilde x}^\mu , y^\alpha\}$, given by the following action
\begin{eqnarray}\label{a.31}
\tilde S &=& \frac{1}{2} \int d\sigma^{+}  d\sigma^{-} ~{\mathbb{{\tilde F}}^{+^{AB}}}(\tilde g , y^{\alpha})~{\tilde R}_{+_{A}}~{\tilde R}_{-_{B}}\nonumber\\
&=&\frac{1}{2} \int d\sigma^{+}  d\sigma^{-}\Big[{{\mathbb{\tilde E}}^{+^{ab}}}(\tilde g , y^{\alpha})~
{\tilde R}_{+_{a}}\;{\tilde R}_{-_{b}}+{\tilde \Phi}^{\hspace{-1mm}+^{(1)^{ a}}}_{~~~~\beta}(\tilde g,y^{\alpha}) {\tilde R}_{+_{a}}\partial_{-} y^{\beta}\nonumber\\
&&~~~~~~~~~~~~~~~~~~~~~~~~~~~~~~~~~+
{\tilde \Phi}^{\hspace{-1mm}+^{(2)^{ b}}}_{\alpha}(\tilde g,y^{\alpha}) \partial_{+} y^{\alpha} ~{\tilde R}_{-_{b}}+{\tilde \Phi}_{_{\alpha \beta}}(y^{\alpha})
\partial_{+} y^{\alpha} \partial_{-} y^{\beta}\Big],
\end{eqnarray}
where ${\tilde R}_{{\pm}_{a}} =
\partial_{_{\pm}} {\tilde x}^\mu {\tilde R}_{_{\mu a}}$ are
the components of the right invariant one-forms on the Lie group ${\tilde {\bf G}}$.
The coupling matrices of the dual sigma model are also determined in a completely analogous
way  as \eqref{a.6} and \eqref{a.13}-\eqref{a.16} such that
by using \eqref{a.29} one relates them to those of the original one by \cite{Klimcik1},  \cite{Tyurin}, \cite{Sfetsos1}, \cite{Sfetsos3}
\begin{eqnarray}
{\mathbb{\tilde E}}^{+}\tiny(\tilde g,y^{\alpha})&=&\Big(E^{+}_{0}\tiny(e,y^{\alpha})+ \tilde \Pi(\tilde g)\Big)^{-1},\label{a.32}\\
{\tilde \Phi}^{+^{(1)}}(\tilde g,y^{\alpha})&=&  {\mathbb{\tilde E}}^{+}\tiny(\tilde g,y^{\alpha})~F^{+^{(1)}}(e , y^{\alpha}),\label{a.33}\\
{\tilde \Phi}^{+^{(2)}}(\tilde g,y^{\alpha})&=& - F^{+^{(2)}}(e , y^{\alpha}) ~{\mathbb{\tilde E}}^{+}\tiny(\tilde g,y^{\alpha}),\label{a.34}\\
{\tilde \Phi}(y^{\alpha})&=& F(y^{\alpha})-F^{+^{(2)}}(e , y^{\alpha})~{\mathbb{\tilde E}}^{+}\tiny(\tilde g,y^{\alpha})~F^{+^{(1)}}(e , y^{\alpha}).\label{a.35}
\end{eqnarray}
Notice that $\tilde \Pi(\tilde g)$ is defined as \eqref{a.8} by replacing untilded quantities by tilded ones.
Hence, the actions \eqref{a.27} and \eqref{a.31}  correspond to Poisson-Lie T-dual sigma models \cite{Klimcik1}.

\vspace{-2mm}
\section{\large  The Abelian T-dualization of the BTZ black hole solutions}

In this section we give the Abelian T-dualization of the BTZ black hole solutions \cite{Banados} by making use of the
approach of the Poisson-Lie T-duality with the spectators reviewed in the previous section.
Let us now begin this section by reviewing the BTZ black hole solutions \cite{Banados}.
The BTZ black holes  are $2+1$-dimensional  solutions of Einstein's equations
with a negative cosmological constant, $\Lambda$,
\begin{eqnarray}\label{b.1}
R_{_{\Upsilon\Delta}} +(\Lambda - \frac{1}{2} R)~ G_{_{\Upsilon\Delta}} = 0, \quad ~ \Lambda <0,
\end{eqnarray}
where  $R_{_{\Upsilon \Delta}}$ and  $R$ are the respective Ricci tensor and  scalar curvature.
The line element for the  black hole solutions is given by
\begin{eqnarray}\label{b.2}
ds^2  = ({\bf M} - {  r^2\over l^2}) d t^2 -Jdtd\phi  + r^2 d  \phi^2 + \big({  r^2\over l^2} - {\bf M}  +
{J^2\over 4r^2}\big)^{-1} d  r^2, \quad   0 \leq \phi < 2\pi
\end{eqnarray}
where the radius $l$ is related to the cosmological constant by $l = (-\Lambda)^{-1/2}.$
The constants of motion ${\bf M}$ and $J$ are the mass and angular momentum of the BTZ black hole, respectively.
They are appeared due to the time translation symmetry and rotational symmetry of the metric,
corresponding to the killing vectors ${\partial / \partial t}$ and ${\partial/\partial \varphi}$, respectively.
The line element \eqref{b.2} describes a black hole solution with outer and inner horizons
at $r =r_+$ and $r= r_-$, respectively,
\begin{eqnarray}\label{b.3}
r_{\pm }= l \bigl ( {{\bf M}\over 2}\bigr )^{1/2}
\biggl \{1\pm \biggl ( 1- \bigl ( {J\over {\bf M}l} \bigr )^2
\biggr )^{1/2}\biggr \}^{1/2},
\end{eqnarray}
where the mass and angular momentum are related to $r_{\pm }$ by
\begin{eqnarray}\nonumber
{\bf M} = \frac{r_{+}^2 + r_-^2}{l^2},~~~~~~~~~~~J = \frac{ 2 r_{+}  r_-}{l}.
\end{eqnarray}
The region $r_+  < r < {\bf M}^{1/2} l$ defines an ergosphere, in which the
asymptotic timelike Killing field $\partial/ {\partial t}$ becomes spacelike.
The solutions with  $- 1 < {\bf M} < 0, \;J=0$  describe  point
particle sources with   naked conical singularities  at
$r=0  $. The  metric with ${\bf M}= -{1 },\; J=0 $ may be recognized as that of ordinary anti-de Sitter space; it
is separated by a mass gap from the ${\bf M}= 0,\; J=0 $.
The vacuum state which is regarded as empty space,  is obtained by letting the horizon size go to zero. This
amounts to letting ${\bf M} \rightarrow 0$, which requires $J \rightarrow 0$. We have to
notice that the metric for the ${\bf M} = J = 0$ black hole is not the same as $AdS_3$ metric which has negative mass ${\bf M}=-1$.
Locally they are equivalent since there is locally only one constant curvature metric in three dimensions.
However they are inequivalent globally.

In Ref. \cite{Horowitz2}, first the BTZ black hole solutions have been considered  in the context of the
low energy approximation, then as the exact conformal field theory.  In three
dimensions, the low energy string effective action is given by
\begin{eqnarray}\label{b.4}
S_{_{eff}}~=~ \int d^3 \Phi ~ \sqrt{-G}~ e^{-2\phi} \Big[R+ 4({\nabla} \phi)^2-\frac{1}{12} H_{_{\Upsilon \Delta \Xi}} H^{^{\Upsilon \Delta \Xi}}-4\Lambda\Big],
\end{eqnarray}
where $G=\det G_{_{\Upsilon\Delta}}$. $H_{_{\Upsilon \Delta \Xi}}$ are the components of the torsion
of the antisymmetric field $B$
and are defined by $H_{_{\Upsilon \Delta \Xi}}=\partial_{_\Upsilon} B_{_{\Delta \Xi}}+
\partial_{_{\Delta}} B_{_{\Xi \Upsilon}}+\partial_{_{\Xi}} B_{_{\Upsilon \Delta}}$, while
${\phi}$ is the dilaton field.
The equations of motion which follow from this action are \cite{callan}
\begin{eqnarray}
0&=&R_{_{\Upsilon \Delta}}-\frac{1}{4} H_{_{\Upsilon \Omega \Xi}}  H^{^{\Omega\Xi}}_{_{~~\Delta}}+2{\nabla}_{_{\Upsilon}}
{\nabla}_{_{\Delta}} \phi,\nonumber\\
0&=&-\frac{1}{2}{\nabla}^{^{\Xi}} H_{_{\Xi \Upsilon \Delta}} + H_{_{\Upsilon \Delta}}^{~\;^{\Xi}} ~ {\nabla}_{_{\Xi}} \phi,\label{b.5}\\
0&=& R + 4 {\nabla}^2 \phi -4 ({\nabla} \phi)^2-\frac{1}{12} H_{_{\Upsilon \Delta \Xi}} H^{^{\Upsilon \Delta \Xi}}-4\Lambda.\nonumber
\end{eqnarray}
These equations are known  as the vanishing of the one-loop beta-functions equations. Notice that
the string effective action is connected to the two-dimensional non-linear sigma
model through these equations \cite{callan}.

In order to obtain an exact solution to string theory, one must modify the BTZ black hole solutions
by  adding an antisymmetric tensor field $H_{_{\Upsilon \Delta \Xi}}$ proportional to the volume form $\epsilon_{_{\Upsilon \Delta \Xi}}$.
It has been shown \cite{Horowitz2} that any solution to three-dimensional
general relativity with negative cosmological constant is a solution to low energy string theory with
$\phi =0$, $H_{_{\Upsilon \Delta \Xi}} = 2\epsilon_{_{\Upsilon \Delta \Xi}}/l$ and $\Lambda = -1/l^2$.
In particular it was shown \cite{Horowitz2} that the two parameter
family of black holes \eqref{b.2} along with
\begin{eqnarray}
B_{\varphi t} = \frac{r^2}{l},~~~~~~~~~~~\phi =0,\label{b.6}
\end{eqnarray}
satisfy the equations of motion \eqref{b.5}. Then, it was obtained \cite{Horowitz2}
the dual of this solution by Buscher's duality transformations \cite{Buscher1}. Abelian duality is a
well known symmetry of string theory that maps any solution of the low energy string
equations of motion \eqref{b.5} with a translational symmetry to another solution.

As a concluding remark for this section we first explicitly reobtain  Buscher's duality transformations \cite{Buscher1} using the approach of the
Poisson-Lie T-duality with the spectators. Then, in this way, we obtain
the Abelian T-dualization of the BTZ black hole solutions given by \eqref{b.2}.  In this case,
the Lie groups $\bf G$ and $\tilde {\bf G}$ are Abelian. Therefore,
both structures $\Pi_{ab}(g)$ and ${\tilde \Pi}^{ab} {(\tilde g)}$ are vanished. Consequently, the background matrices
${\mathbb{F}_{_{AB}}^{+}}(g,y^{\alpha})$ and ${\mathbb{{\tilde F}}^{+^{AB}}}(\tilde g , y^{\alpha})$ are equaled with
$E^{+}_{_{AB}}(e,y^{\alpha})$ and ${{\tilde E}^{+^{AB}}}({\tilde e} , y^{\alpha})$, respectively. Thus, the action
\eqref{a.27} in the presence of the dilaton field $\phi$  is written as
\begin{eqnarray}\label{b.7}
S &=&\frac{1}{2} \int d\sigma^{+}  d\sigma^{-}\Big[{{E}_{_{AB}}^{+}}(e,y^{\alpha})~
\delta_{_{\Upsilon}}^{^{~A}}\;\delta_{_{\Lambda}}^{^{~B}} ~ \partial_{+}\Phi^{^{\Upsilon}}~\partial_{-}\Phi^{^{\Lambda}}-\frac{1}{4} R^{(2)} \phi\Big],
\end{eqnarray}
where $R^{(2)}$ is two-dimensional scalar curvature in the worldsheet. Comparing the sigma model \eqref{b.7} and \eqref{a.3} we find that
${{E}_{_{AB}}^{+}}(e,y^{\alpha})~
\delta_{_{\Upsilon}}^{^{~A}}\;\delta_{_{\Lambda}}^{^{~B}} = G_{_{\Upsilon \Lambda}} + B_{_{\Upsilon \Lambda}}$.
Let us assume that the sigma model \eqref{b.7} has an Abelian isometry represented by a translation in a coordinate $x^0$ in the target space.
Notice that in the coordinates
$\Phi^{^{\Upsilon}}=\{x^0 , x^\mu\}$, $\mu = 1,...,d-1$ adapted to the isometry, the metric, torsion and dilaton field are independent
of ${x^0}$.
If we choose the matrix $E_{_{AB}}^+(e,y^\alpha)$ as
\begin{eqnarray}\label{b.8}
E_{_{AB}}^+(e,y^\alpha)\;=\;\left( \begin{array}{cc}
                     G_{_{00}} & G_{_{0 \nu}}+B_{_{0 \nu}}\\

                     G_{_{\mu 0}}+B_{_{\mu 0}} & G_{_{\mu \nu}}+B_{_{\mu \nu}}
                      \end{array} \right),
\end{eqnarray}
then from relations \eqref{a.29} and \eqref{a.30}
one immediately  gets the following Buscher's duality transformations \cite{Buscher1}
\begin{eqnarray}\label{b.9}
\left( \begin{array}{cc}
                    \tilde G_{_{00}} & \tilde G_{_{0 \nu}}+\tilde B_{_{0 \nu}}\\

                    \tilde G_{_{\mu 0}}+\tilde B_{_{\mu 0}} &\tilde G_{_{\mu \nu}}+\tilde B_{_{\mu \nu}}
                      \end{array} \right)=\left( \begin{array}{cc}
                     \frac{1}{G_{00}} &  \frac{G_{_{0 \nu}}+B_{0 \nu}}{G_{_{00}}}\\

                    - \frac{G_{_{\mu 0}}+B_{_{\mu 0}}}{G_{_{00}}} & G_{_{\mu \nu}}+B_{_{\mu \nu}}
                    -\frac{(G_{_{\mu 0}}+B_{_{\mu 0}})(G_{_{0 \nu}}+B_{_{0 \nu}})}{G_{_{00}}}
                      \end{array} \right).
\end{eqnarray}
It has been shown \cite{Buscher1}, \cite{Buscher2} that for preserving conformal invariance, at least to one-loop order,
the dilaton field has to transform as
\begin{eqnarray}\label{b.10}
{\tilde \phi}~=~\phi-\frac{1}{2} \ln G_{_{00}}.
\end{eqnarray}
This shift of the dilaton field is due to the duality transformation receiving corrections from the jacobian that
comes from integrating out the gauge fields \cite{Buscher1}, \cite{Buscher2}.
In the Poisson-Lie T-duality case, dilaton shifts in both models have been
obtained by quantum considerations based on a regularization of a functional
determinant in a path integral formulation of Poisson-Lie duality by
incorporating spectator fields \cite{Tyurin} (see, also, \cite{N.Mohammedi})
\begin{eqnarray}
\phi &=&\phi_{_{0}}(y^{\alpha})+\frac{1}{2}\ln \big(\det {\mathbb{E}}^{+}\tiny(g,y^{\alpha})\big) -\frac{1}{2} \ln \big(\det {E_0^+(e , y^{\alpha})}\big),\label{b.10.1}\\
{\tilde \phi} &=&\phi_{_{0}}(y^{\alpha})+\frac{1}{2}\ln \big(\det {\mathbb{\tilde E}}^{+}\tiny(\tilde g,y^{\alpha})\big).\label{b.10.2}
\end{eqnarray}
Now one can use the above approach to obtain the Abelian T-dual solutions with the BTZ black hole solutions.
Here the target space ${\cal M} \approx O \times \bf G$ is defined by the coordinates $\{\varphi, r, t\}$,
where the Lie group $\bf G$ should be considered to be $U(1)$. It is more convenient to write the action of
sigma model corresponding to the BTZ metric \eqref{b.2} and the solutions \eqref{b.6}. It is then read to be in the following form
\begin{eqnarray}\label{b.11}
S &=&\frac{1}{2} \int d\sigma^{+}  d\sigma^{-}\Big[r^2~\partial_{+}\varphi~\partial_{-}\varphi
+\Big(\frac{r^2}{l^2}-{\bf M}+\frac{J^2}{4r^2}\Big)^{-1}~ \partial_{+}r ~\partial_{-}r\nonumber\\
&&~~~~~~~~~~~~~+\big({\bf M}-\frac{r^2}{l^2}\big)~\partial_{+}t ~\partial_{-}t -
\big(\frac{J}{2} -\frac{r^2}{l}\big)~\partial_{+}\varphi ~\partial_{-}t -\big(\frac{J}{2} +\frac{r^2}{l}\big)~\partial_{+}t ~\partial_{-}\varphi\Big].
\end{eqnarray}
On the other hand, since the background described by the action \eqref{b.11} depend on the ${r}$ coordinate only,
from the Poisson-Lie T-duality standpoint the Lie group $\bf G$ can not be parametrized by this coordinate.
However, we assume that the $\bf G$ is parametrized by the ${\varphi}$ coordinate while the coordinates of orbit $O$
are represented by the ${r}$ and ${t}$.
Comparing the action \eqref{b.11} and \eqref{b.7} the matrix $E_{_{AB}}^+(e,y^\alpha)$ is derived as
\begin{eqnarray}\label{b.12}
E_{_{AB}}^+(e,y^\alpha)\;=\;\left( \begin{array}{ccc}
                    r^2 & 0 & -\frac{J}{2} +\frac{r^2}{l}\\
                   0 & \Big(\frac{r^2}{l^2}-{\bf M}+\frac{J^2}{4r^2}\Big)^{-1} & 0\\
                     -\frac{J}{2} -\frac{r^2}{l} & 0  & {\bf M}-\frac{r^2}{l^2}
                      \end{array} \right).
\end{eqnarray}
Considering \eqref{b.12} in the form \eqref{a.6} and using the fact that ${\bf G} =U(1)$ we find that $E^{+}_{0\;ab}\tiny(e,y^{\alpha})=r^2 $ and
{\small
\begin{eqnarray}\label{b.13}
F^{+^{(1)}}_{a \beta}=\left( \begin{array}{cc}
                    0 & -\frac{J}{2} +\frac{r^2}{l}
                      \end{array} \right),~~~
                      F^{+^{(2)}}_{\alpha b}=\left( \begin{array}{c}
                      0\\
                       -\frac{J}{2} -\frac{r^2}{l}
                      \end{array} \right),~~
                      F_{\alpha \beta}=\left( \begin{array}{cc}
                     \Big(\frac{r^2}{l^2}-{\bf M}+\frac{J^2}{4r^2}\Big)^{-1} & 0\\
                     0  & {\bf M}-\frac{r^2}{l^2}
                      \end{array} \right).
\end{eqnarray}
}
The dual target space $\tilde {\cal M}  \approx O \times {\tilde {\bf G}}$ is defined by the coordinates
$\{\tilde \varphi,  r,  t\}$. Obviously, in the standard Abelian case ${\bf D} = U(1) \times U(1)$, i.e.,  ${\tilde {\bf G}}$ is
also identical to $U(1)$ and is parametrized by the $\tilde \varphi$ coordinate.
Finally, utilizing the equations \eqref{a.32}-\eqref{a.35} together with \eqref{a.31} the dual sigma model is read off to be of the form
{\small\begin{eqnarray}\label{b.14}
\tilde S &=&\frac{1}{2} \int d\sigma^{+}  d\sigma^{-}\Big[\frac{1}{r^2}~\partial_{+}{\tilde \varphi}~\partial_{-}{\tilde \varphi}
+\Big(\frac{r^2}{l^2}-{\bf M}+\frac{J^2}{4r^2}\Big)^{-1}~ \partial_{+}{ r} ~\partial_{-}{ r}\nonumber\\
&&~~~~+\big({\bf M}-\frac{J^2}{4r^2}\big)~\partial_{+}{ t} ~\partial_{-}{t} +\big(-\frac{J}{2r^2} +\frac{1}{l}\big)~
\partial_{+}{\tilde \varphi} ~\partial_{-}{t}+
\big(\frac{J}{2r^2} +\frac{1}{l}\big)~\partial_{+}{t} ~\partial_{-}{\tilde \varphi}-\frac{1}{4} R^{(2)}~{\tilde \phi}\Big],
\end{eqnarray} }
where the new dilaton  ${\tilde \phi}$ at the last term of the action can be calculated by \eqref{b.10} to be of the form
\begin{eqnarray}\label{b.15}
{\tilde \phi} ~=~-\ln r.
\end{eqnarray}
In this way the dual geometry is given by
\begin{eqnarray}
{d {\tilde s}}^2 &=&\big({\bf M}-\frac{J^2}{4r^2}\big) d t^2+ \Big(\frac{r^2}{l^2}-{\bf M}+\frac{J^2}{4r^2}\Big)^{-1} d r^2+ \frac{2}{l}
~d t~ d \tilde \varphi+   \frac{1}{r^2} d {\tilde \varphi}^2,\label{b.16}\\
{\tilde B} &=& - \frac{J}{2r^2}~d {\tilde \varphi} \wedge d t.\label{b.17}
\end{eqnarray}
The solutions \eqref{b.15}-\eqref{b.17} have been obtained in Ref. \cite{Horowitz2} by using  Buscher's duality transformations.
The three-dimensional charged black string solutions can be obtained from the dual solutions
\eqref{b.15}-\eqref{b.17} by making the coordinate transformation $t=l(r_+^2 - r_-^2)^{^{\frac{-1}{2}}} (\hat{x}-\hat{t})$,~
${\tilde \varphi}=l(r_+^2 - r_-^2)^{^{\frac{-1}{2}}} (r_+^2 \hat{t}- r_-^2 \hat{x})$ and $r^2 = l \hat{r}$ as follows \cite{Horowitz2}
\begin{eqnarray}
{d {\tilde s}}^2 &=&-\big(1-\frac{\mathbb{M}}{\hat{r}}\big) d \hat{t}^2+ \Big(1-\frac{{\mathbb{Q}}^2}{{\mathbb{M}} \hat{r}}\Big) d \hat{x}^2+
\big(1-\frac{\mathbb{M}}{\hat{r}}\big)^{^{-1}}  \Big(1-\frac{{\mathbb{Q}}^2}{{\mathbb{M}} \hat{r}}\Big)^{^{-1}}~ \frac{l^2 d \hat{r}^2}{4 \hat{r}^2},\label{b.20}\\
{\tilde B} &=& \frac{\mathbb{Q}}{\hat{r}}~d {\hat{x}} \wedge d \hat{t},\label{b.21}\\
{\tilde \phi} & = & -\frac{1}{2} \ln (l \hat{r}),\label{b.22}
\end{eqnarray}
where ${{\mathbb{M}}} = {r_+^2}/l$ and ${{\mathbb{Q}}} = {J}/2$ are the respective the mass and charge of the black string.
For large $\hat{r}$ the metric is asymptotically
flat. Thus, it is important to notice that the Abelian T-duality  transformation changes the asymptotic behavior from
$AdS_3$ to flat. In addition, as it can be seen from the above results,  for asymptotically flat solutions of
three-dimensional low energy string theory, the duality of conserved quantities defined on
asymptotic region is given in such a way that a mass is unchanged, while an axion charge
and an angular momentum are interchanged each other \cite{Horowitz2}.
In the next section we give the main goal of the paper. We will study the non-Abelian T-dulaization  of
the BTZ black hole solution in such a way that we shall show that the dual model describes a new three-dimensional charged
black string.

\section{\large The BTZ black hole vacuum solution from  Poisson-Lie T-dual sigma models with the spectators }

In this section we explicitly construct  a pair of Poisson-Lie T-dual sigma models on $2+1$-dimensional
manifolds ${\cal M}$ and $\tilde {\cal M}$ as the target spaces.
The original model is built on the manifold ${\cal M} \approx O \times {\bf G}$, where ${\bf G}=A_2$ as a
two-dimensional real non-Abelian Lie group acts freely on it while $O$ is the orbit of $\bf G$ in ${\cal M}$ with one
spectator $y^\alpha =\{y\}$.
The target space of the dual model is the manifold $\tilde {\cal M} \approx O \times \tilde {\bf G}$
with two-dimensional real Abelian Lie group ${\tilde {\bf G}}=2A_1$ acting freely on it. We shall show
the original model is canonically equivalent to the $SL(2,\mathbb{R})$ WZW model  for a given value of the background parameters.
In particular, by a convenient coordinate transformation we show that the model
describes a string propagating in a spacetime with the BTZ black hole vacuum metric.

\vspace{-2mm}
\subsection{\small The original model as the $SL(2,\mathbb{R})$ WZW model and BTZ black hole vacuum solution}

As mentioned above the original model is built on a $2+1$-dimensional manifold ${\cal M} \approx O \times {\bf G}$ in which ${\bf G}$ is
two-dimensional real non-Abelian Lie group whose Lie algebra is denoted by ${\cal A}_2$.
According to Section 2,  having  Drinfeld doubles we can construct the Poisson-Lie T-dual sigma models on them.
The Lie algebra of the Drinfeld double $({\cal A}_2 , 2{\cal A}_1)$ is defined by the following non-zero Lie brackets \cite{Hlavaty1}
\begin{eqnarray}\label{d.1}
[T_1 , T_2]~=~T_2,~~~~~[T_1 ~, ~{\tilde T}^2]=-{\tilde T}^2,~~~~~[T_2 ~, ~{\tilde T}^2]={\tilde T}^1.
\end{eqnarray}
We note that the  Lie algebra $({\cal A}_2 , 2{\cal A}_1)$ is isomorphic to the oscillator Lie algebra $h_4$  \cite{eghbali11}.
According to action \eqref{a.27} to construct the original model we need the components of right invariant one-forms
$R_{\pm}^a$ on the Lie group $A_2$. To this end, we parametrize an element of $A_2$ as
\begin{eqnarray}\label{d.2}
g~=~e^{x_1 T_1}~e^{x_2 T_2},
\end{eqnarray}
where $x_1$ and $x_2$ are the coordinates of the Lie group $A_2$. Then $R_{\pm}^a$'s are derived to be of the form
\begin{eqnarray}\label{d.3}
R_{\pm}^1~=~ \partial_{\pm} x_1,~~~~~~~~~~~~R_{\pm}^2~=~ e^{x_1}~\partial_{\pm} x_2.
\end{eqnarray}
We note that since the dual Lie group have been considered to be Abelian,
hence, by using \eqref{a.7} and \eqref{a.8}  it follows  that the
$\Pi_{ab}(g)$ is vanished. In addition to \eqref{d.3}, we need to determine the couplings $\mathbb{E}^{+}_{ab}\tiny(g , y^\alpha)$,
$\Phi^{+^{(1)}}_{a \beta}\tiny{(g , y^\alpha)}$ and $\Phi^{+^{(2)}}_{\alpha b}\tiny{(g , y^\alpha)}$. By a {\it convenient choice} of
the matrices
$E^{+}_{0\;ab}\tiny(e , y^\alpha)$, $F^{+^{(1)}}_{a \beta}\tiny(e , y^\alpha)$ and $F^{+^{(2)}}_{\alpha b}\tiny(e , y^\alpha)$ as
{\small{\begin{eqnarray}\label{d.4}
E^{+}_{0\;ab}(e , y^\alpha)=\left( \begin{array}{cc}
                    0 & \frac{1}{2}e^{-2y}\\
                    \frac{1}{2}e^{-2y} & 0
                      \end{array} \right),~~~
F^{+^{(1)}}_{a \beta}(e , y^\alpha)=\left( \begin{array}{c}
                             0\\
                            -e^{-2y}
                      \end{array} \right),~~~
F^{+^{(2)}}_{\alpha b}(e , y^\alpha)=\left( \begin{array}{cc}
                    0 & e^{-2y}
                      \end{array} \right),
\end{eqnarray}}}
and $F_{\alpha\beta} (y^\alpha)=k$ ($k$ is a non-zero real constant),
and then with the help of relations \eqref{a.13}-\eqref{a.16} one can get the required couplings. Finally
by inserting these into action \eqref{a.27} the original sigma model is found to be of the form
\begin{eqnarray}
S &=& \frac{1}{2} \int d \sigma^+ d \sigma^-~\Big[k \partial_+ y \partial_- y  + \frac{1}{2} e^{x_1 - 2y} (\partial_+ x_1 \partial_- x_2
+\partial_+ x_2 \partial_- x_1)\nonumber\\
&&~~~~~~~~~~~~~~~~~~~~~~~~~~~~~~~~~~~~~~-e^{x_1 - 2y}(\partial_+ x_2 \partial_- y- \partial_+ y \partial_- x_2) \Big].\label{d.5}
\end{eqnarray}
By identifying action (\ref{d.5}) with the sigma model
of the form (\ref{a.3}) one can read off the background matrix.
The space-time  metric and the antisymmetric tensor field corresponding to the action (\ref{d.5}) can be written as
\begin{eqnarray}
d {s}^2 &=&k~ d y^2 + e^{x_1 - 2y} d x_1~ d x_2,\label{d.6.1}\\
{B} &=& - e^{x_1 - 2y}~d x_2 \wedge d y.\label{d.6.2}
\end{eqnarray}
For  the metric \eqref{d.6.1} one can find that $R_{_{\Upsilon \Lambda}} =-({2}/{k}) {G}_{_{\Upsilon\Lambda}}$ and then $R=-{6}/{k}$.
This shows that the metric \eqref{d.6.1} with $k>0$ describes an anti-de Sitter space while for $k<0$ we have a de Sitter space.
Considering antisymmetric tensor field ${B}$ of \eqref{d.6.2} one quickly finds that the only non-zero component of
$H$ is $H_{x_{_1} x_{_2} y} = -e^{x_1 - 2y}$, then, it follows that $H_{_{\Upsilon \Delta \Xi}} H^{^{\Upsilon \Delta \Xi}} ={-24}/{k}$.
Inserting the above results in the vanishing of the one-loop beta-functions equations \eqref{b.5},
the conformal invariance conditions up to one-loop order are satisfied with $\Lambda =-{1}/{k}$ and a constant dilaton field
which satisfies the equation \eqref{b.10.1}.

On the one hand, action \eqref{d.5} can be  simplified by making the following new
coordinates
\begin{eqnarray}\label{d.7}
e^{x_1} ~=~ \theta_+,~~~~~~~x_2~=~\theta_-,~~~~~~~~y~=~ \gamma,
\end{eqnarray}
then\footnote{Under the transformation \eqref{d.7}
the action \eqref{d.5} turns into
\begin{eqnarray}
\bar{ S} = \frac{1}{2} \int d \sigma^+ d \sigma^-~\Big[k \partial_+ \gamma \partial_- \gamma  + \frac{1}{2} e^{-2\gamma} (\partial_+ \theta_+ \partial_- \theta_-
+\partial_+ \theta_- \partial_- \theta_+)- \theta_+ e^{-2\gamma}(\partial_+ \theta_- \partial_- \gamma- \partial_+ \gamma \partial_- \theta_-) \Big].\label{d.5.5.1}
\end{eqnarray}
By calculating the momentums corresponding to the actions \eqref{d.5} and \eqref{d.5.5.1} ($ P_{_{\Lambda}}$ and ${\bar P}_{_{\Lambda}}$, respectively) and  by making use of
the transformation \eqref{d.7} we get
the transformation between momentums.
Then, by considering the basic equal-time Poisson brackets for the pair of canonical variables
$(\Phi^{^{\Upsilon}} , P_{_{\Lambda}})$  one can show that
the equal-time Poisson brackets  for the pair of canonical variables
$({\bar \Phi}^{^{\Upsilon}} , {\bar P}_{_{\Lambda}})$ are also preserved. Furthermore, we can obtain the Hamiltonians corresponding to
the actions \eqref{d.5} and \eqref{d.5.5.1}, and then show that they are identical together, that is, under the transformation \eqref{d.7}
one goes from ${\bar {\cal H}}({\bar {\Phi}})$ to ${\cal H}({{\Phi}})$ and vice versa, hence, proving that the transformation \eqref{d.7}
 is indeed a canonical transformation.
}, by making use of the integrating by parts over the last two terms of the action \eqref{d.5.5.1} we get
\begin{eqnarray}\label{d.8}
S ~=~ \frac{1}{2} \int d \sigma^+ d \sigma^- \Big[k \partial_+ \gamma \partial_- \gamma
+ e^{-2 \gamma}  \partial_+ \theta_+ \partial_- \theta_-\Big].
\end{eqnarray}
By setting $k=1$ the above action precisely becomes the $SL(2,\mathbb{R})$ WZW model \cite{Alekseev1}. On the other hand,
to better understand of action \eqref{d.5} we diagonalize the metric obtained by this action.
Let
\begin{eqnarray}
e^{x_1} ~=~ \frac{1}{l}({t}- l {\varphi}),~~~~~~
x_2    ~=~ -l({t}+ l {\varphi}),~~~~~~~~
e^{y}    ~=~ \frac{l}{{r}},\label{d.9}
\end{eqnarray}
where the radius $l$ introduced in Section 3. Then the action \eqref{d.5}  turns into
\begin{eqnarray}
{S} ~=~ \frac{1}{2} \int d \sigma^+ d \sigma^-\hspace{-6mm}&&\Big[-\frac{{{r}}^2}{l^2} \partial_+ {t} \partial_- {t}
+ {{r}}^2  \partial_+ {\varphi} \partial_- {\varphi}+ \frac{l^2}{{{r}}^2} \partial_+ {r} \partial_- {r}\nonumber\\
&&-\frac{{{r}}}{l^2}({t}- l {\varphi}) (\partial_+ {t} \partial_- {r} - \partial_+ {r} \partial_- {t})
-\frac{{{r}}}{l}({t}- l {\varphi}) (\partial_+ {\varphi} \partial_- {r} - \partial_+ {r} \partial_- {\varphi}) \Big],\label{d.10}
\end{eqnarray}
where we have, here, set $k=l^2$. This describes a string propagating in a space-time with the BTZ black hole vacuum metric with
${\bf M} = J = 0$
\begin{eqnarray}\label{d.11}
d {s}^2 ~=~ -\frac{{{r}}^2}{l^2} d {t}^2
+ {{r}}^2 d {\varphi}^2 + \frac{l^2}{{{r}}^2} d {r} ^2,
\end{eqnarray}
and an antisymmetric tensor field
\begin{eqnarray}\label{d.12}
{B} ~=~ -\frac{{{r}}}{l}({t}- l {\varphi}) \Big(d {\varphi} \wedge d {r} + \frac{1}{l} d {t} \wedge d {r}\Big).
\end{eqnarray}
Since the  black hole metric with ${\bf M} = J = 0$ is locally equivalent to the $AdS_3$ metric, so, for the metric \eqref{d.11} one immediately finds that
$R_{_{\Upsilon \Lambda}} =-({2}/{l^2}) {G}_{_{\Upsilon\Lambda}}$ and $R=-{6}/{l^2}$.
The only non-zero component of antisymmetric field strength corresponding to the $B$-field \eqref{d.12} is
$H_{_{t r \varphi}}=2{r}/{l}$; consequently $H_{_{\Upsilon \Delta \Xi}} H^{^{\Upsilon \Delta \Xi}} ={-24}/{l^2}$.
Putting these pieces together, one verifies the  equations \eqref{b.5}
with $\Lambda = -{1}/{l^2}$ and a constant dilaton field.
Therefore, it follows that the conformal invariance of actions
\eqref{d.5} and \eqref{d.10} is, under the transformation \eqref{d.9}, preserved.
Notice that the metric \eqref{d.11} has no horizon and no curvature singularity. Indeed,
this solution is everywhere regular including $r=0$. In fact, this was expected since generally the (anti-)de
Sitter space-time is a maximally symmetric space with a constant curvature.
The resulting space-time is completely nonsingular. Consider now two killing vectors
$\partial/{\partial t}$ and $\partial/{\partial {\varphi}}$ corresponding to the time translational
and the rotational isometries of the metric \eqref{d.11}, respectively. The killing field
$\partial/{\partial t}$ becomes null at $r=0$ and it is timelike for the whole range $r>0$, while
the killing field   $\partial/{\partial {\varphi}}$ is everywhere spacelike except for  $r=0$.
The global structure associated with the metric \eqref{d.11}  including the Kruskal and Penrose
diagrams has been discussed in Ref. \cite{Banados1}.


\subsection{\small The dual model as a three-dimensional charged black string}
As mentioned at the first of this section, the dual model is constructed on a $2+1$-dimensional manifold
$\tilde {\cal M} \approx O \times \tilde {\bf G}$
with two-dimensional Abelian Lie group ${\tilde {\bf G}}=2A_1$ acting freely on it.
In the same way to construct out the dual sigma model we parametrize the corresponding Lie
group (Abelian Lie group $2A_1$) with  coordinates ${\tilde x}^\mu = \{{\tilde x}_1 , {\tilde x}_2\}$ so that its elements
are defined as \eqref{d.2} by replacing untilded quantities by tilded ones. Hence the components of the right invariant one-forms
on the Lie group $2A_1$ are simply calculated to be ${\tilde R}_{\pm_{a}}=\partial_{\pm} {{\tilde x}_a}$. Utilizing relation
\eqref{a.8} for untilded quantities we get
\begin{eqnarray}\label{c.1}
{\tilde \Pi}^{ab}(\tilde g)\;=\;\left( \begin{array}{cc}
                     0 & -{\tilde x}_2\\

                     {\tilde x}_2 & 0
                      \end{array} \right).
\end{eqnarray}
Inserting \eqref{d.4} and \eqref{c.1} into equations \eqref{a.32}-\eqref{a.35}  the dual couplings are obtained to be of the form
{\small{\begin{eqnarray}\label{c.2}
\mathbb{\tilde E}^{+^{ab}}=\left( \begin{array}{cc}
                    0 & \frac{1}{{\tilde x}_2+\frac{1}{2}e^{-2y}}\\
                    \frac{1}{-{\tilde x}_2+\frac{1}{2}e^{-2y}} & 0
                      \end{array} \right),~~~~
{\tilde \Phi}^{\hspace{-1mm}+^{(1)^{ a}}}_{~~~~\beta}=\left( \begin{array}{c}
                             \frac{-e^{-2y}}{{\tilde x}_2+\frac{1}{2}e^{-2y}}\\
                            0
                      \end{array} \right),~~~~~
{{\tilde \Phi}^{+^{{(2)}^{b}}}}_{~\alpha}=\left( \begin{array}{cc}
                     \frac{-e^{-2y}}{-{\tilde x}_2+\frac{1}{2}e^{-2y}} & 0
                      \end{array} \right),
\end{eqnarray}}}
and ${\tilde \Phi}_{_{\alpha \beta}} = k$. Putting these pieces together into \eqref{a.31}, the action of dual model is worked out to be
\begin{eqnarray}
{\tilde S} &=& \frac{1}{2} \int d \sigma^+ d \sigma^-\Big\{k~\partial_+y~\partial_-y +\frac{1}{{\Delta}}\Big[(\frac{1}{2}e^{-2y}-{\tilde x}_2) \partial_+{{\tilde x}_1}~
\partial_-{{\tilde x}_2}+(\frac{1}{2}e^{-2y}+{\tilde x}_2) \partial_+{{\tilde x}_2}~
\partial_-{{\tilde x}_1}\nonumber\\
&&~~~~~~~~~~~~~~~~~~~-e^{-2y}(\frac{1}{2}e^{-2y}-{\tilde x}_2) \partial_+{{\tilde x}_1}~\partial_-y-e^{-2y}(\frac{1}{2}e^{-2y}+{\tilde x}_2)
\partial_+y~\partial_-{{\tilde x}_1}\Big]\Big\},\label{c.3}
\end{eqnarray}
where ${\Delta} = \frac{1}{4}e^{-4y}-{{\tilde x}_2}^2$. Comparing the above action with the sigma model action of the form \eqref{a.3},
the corresponding metric and tensor field $\tilde B$ take the following forms
\begin{eqnarray}
{d {\tilde s}}^2 &=&k ~d y^2+ \frac{e^{-2y}}{{\Delta}}\Big(d {{\tilde x}_1}~d {{\tilde x}_2}-
e^{-2y}~ d {{\tilde x}_1}~d y\Big),\label{c.4}\\
{\tilde B} &=& - \frac{{{\tilde x}_2}}{{\Delta}}\Big(d {{\tilde x}_1} \wedge d {{\tilde x}_2}-
e^{-2y}~ d {{\tilde x}_1}  \wedge d y\Big).\label{c.5}
\end{eqnarray}
As mentioned in Section 2 the spectator fields do not participate in the Poisson-Lie T-duality transformations.
Therefore, as it can be seen from  the  metrics \eqref{d.6.1} and \eqref{c.4} this duality changes the group part of the metrics, while
leaving the $y$ component invariant.
The metric components \eqref{c.4} are ill defined at the regions ${\tilde x}_2 = \frac{1}{2} e^{-2y}$ and
${\tilde x}_2 = - \frac{1}{2} e^{-2y}$. We can test whether  there are true singularities
by calculating the scalar curvature, which is
\begin{eqnarray}
{\tilde R} ~=~ - \frac{2(11 e^{-4y} +28 {\tilde x}_2 e^{-2y} +12 {{\tilde x}_2}^2)}{k(e^{-2y} - 2{\tilde x}_2)^2}.\label{c.6}
\end{eqnarray}
Thus  ${\tilde x}_2 = \frac{1}{2} e^{-2y}$ is a true curvature singularity, while the difficult at the
${\tilde x}_2 = - \frac{1}{2} e^{-2y}$ can be removed by an appropriate change of coordinates.
As shown in subsection 4.1 the original model \eqref{d.5} is canonically equivalent to the $SL(2 , \mathbb{R})$ WZW model.
Therefore, it should be canformally invariant. To check
the conformal invariance of the dual model \eqref{c.3}
we look at vanishing of the one-loop beta-functions equations \eqref{b.5}.
Given a $\tilde B$-field with \eqref{c.5} we find that the only non-zero component of
$\tilde H$ is ${\tilde H}_{_{{\tilde x}_1 {\tilde x}_2 y}}=- 4 e^{-2y}/(e^{-2y} - 2{\tilde x}_2)^2$.
Thus, the first two equations of \eqref{b.5} are satisfied by the new dilaton field
\begin{eqnarray}
{\tilde \phi} ~=~ a +\frac{1}{2} \ln \Big(\frac{2{\tilde x}_2 + e^{-2y}}{2{\tilde x}_2 - e^{-2y}}\Big),\label{c.8}
\end{eqnarray}
where $a$ is an arbitrary constant. This additive constant plays an important role, as we will see
later. Thus, the dilatonic contribution  in  \eqref{b.5} vanishes if the cosmological constant of the dual theory does leave
invariant, that is, ${\tilde \Lambda} = -1/k$. It is also important to note that the dilaton field \eqref{c.8} is well behaved
for the ranges ${\tilde x}_2 +\frac{1}{2} e^{-2y} <0$ and ${\tilde x}_2 - \frac{1}{2} e^{-2y}>0$.


As shown, the metric \eqref{d.11} of the original model  is the BTZ black hole vacuum solution which is locally equivalent to the $AdS_3$ metric.
To continue,  we shall show that the dual solution represents a three-dimensional charged black string which is stationary
and asymptotically flat. In order to enhancing and clarifying the structure of the dual space-time, horizon, singularity and also
determining the asymptotic nature of one, we first write the solutions \eqref{c.4}, \eqref{c.5} and
\eqref{c.8} in the coordinate base $\{{t}, {x}, {r}\}$. Furthermore, since we want to discuss
the dual solution to the BTZ black hole vacuum solution, we must here consider $k=l^2$.
In the following, we separately discuss the solutions for the ranges ${\tilde x}_2 +\frac{1}{2} e^{-2y} <0$
and ${\tilde x}_2 - \frac{1}{2} e^{-2y}>0$.

\bigskip
$\bullet$~ {\bf \small The solution corresponding to  the range ${\bf {\tilde x}_2 +\frac{1}{2} e^{-2y} <0}$ }

\smallskip
In this case we consider ${\tilde x}_2 +\frac{1}{2} e^{-2y} = - e^T$ for  $T \in \Re$. Then, we introduce
the following coordinate transformation
\begin{eqnarray}
{{\tilde x}_1} =  U +\frac{l^2}{2} (W+e^W),~~~~~{{\tilde x}_2} = -e^T (1+\frac{e^{-W}}{2}),~~~~~~
{{\tilde y}} = \frac{1}{2} (W-T),\label{x.1}
\end{eqnarray}
for  $U, W \in \Re$. Under the above transformation, the dual metric \eqref{c.4},
the $\tilde B$-field \eqref{c.5} and the dilaton field \eqref{c.8} are, respectively, turn into
\begin{eqnarray}
{d {\tilde s}}^2 &=&\frac{l^2}{4} ~d W^2+ \frac{l^2}{4} ~d T^2 + \frac{1}{e^W+1} ~d T d U,\label{x.2}\\
{\tilde B} &=&  \big[1-\frac{1}{2(e^W+1)}\big] ~ d U \wedge d T + \frac{l^2}{2} (e^W+\frac{1}{2}) ~d W \wedge d T,\label{x.3}\\
{\tilde \phi} &=& a +  \frac{1}{2} \ln \big(\frac{e^W}{e^W+1}\big),\label{x.4}
\end{eqnarray}
Notice that there is no singularity for the metric \eqref{x.2}.  In fact, this was expected since
the solutions \eqref{c.4}, \eqref{c.5} and \eqref{c.8} are, in this case, defined only for the range ${\tilde x}_2 +\frac{1}{2} e^{-2y} <0$.
As explained above, the true singularity of the metric \eqref{c.4} occurs  at ${\tilde x}_2 = \frac{1}{2} e^{-2y}$,
which this region is located out of the range ${\tilde x}_2 +\frac{1}{2} e^{-2y} <0$.
Let us now consider the transformation $e^W = {1}/{(\hat{r} -1)}$ so that it requires that $1 <\hat{r}< \infty$. In addition to,
we introduce the following linear transformation
\begin{eqnarray}
T =  -\frac{2}{l} (\hat{t} +\frac{\hat{x}}{\sqrt{3}}), ~~~~~~~~ U = l (\hat{t} - \frac{\hat{x}}{\sqrt{3}}),\label{x.5}
\end{eqnarray}
for $\hat{t}, \hat{x} \in \Re$.
By applying the above transformations to the solutions \eqref{x.2}, \eqref{x.3} and \eqref{x.4}, one obtains
the forms of the dual space-time metric, antisymmetric field strength and dilaton field in new coordinate base $\{{\hat{t}}, {\hat{x}}, {\hat{r}}\}$ as
\begin{eqnarray}
{d {\tilde s}}^2 &=& - (1-\frac{2}{\hat{r}}) d \hat{t}^2 +  (1-\frac{2}{3 \hat{r}}) d \hat{x}^2 +
\frac{2}{\sqrt{3}}~ d \hat{t} d \hat{x} +(1-\frac{1}{\hat{r}})^{-2} ~ \frac{l^2 d \hat{r}^2}{4 \hat{r}^2},\label{x.6}\\
{{\tilde H}_{\hat{r} \hat{t} \hat{x}}} &=& \frac{2}{\sqrt{3}~ \hat{r}^2},\label{x.7}\\
{\tilde \phi} &=& a -\frac{1}{2} \ln \hat{r}. \label{x.8}
\end{eqnarray}
We note that this solution is valid only for the range ${\tilde x}_2 +\frac{1}{2} e^{-2y} <0$ or $1 <\hat{r}< \infty$.

\bigskip

$\bullet$~ {\bf \small The solution corresponding to  the range ${\bf {\tilde x}_2 -\frac{1}{2} e^{-2y} >0}$ }

\smallskip
For this case, we have assumed that ${\tilde x}_2 - \frac{1}{2} e^{-2y} = e^T-e^{-2y}$ for which
$T+2y  >0 $. Analogously, by introducing the following transformation
\begin{eqnarray}
{{\tilde x}_1} =  U -\frac{l^2}{2} (e^W-W),~~~~~{{\tilde x}_2} = e^T (1-\frac{e^{-W}}{2}),~~~~~~
{{\tilde y}} = \frac{1}{2} (W-T),\label{x.9}
\end{eqnarray}
we obtain
\begin{eqnarray}
{d {\tilde s}}^2 &=&\frac{l^2}{4} ~d W^2+ \frac{l^2}{4} ~d T^2 - \frac{1}{e^W-1} ~d T d U,\label{x.10}\\
{\tilde B} &=&  \big[1+\frac{1}{2(e^W-1)}\big] ~ d U \wedge d T - \frac{l^2}{2} (e^W-\frac{1}{2}) ~d W \wedge d T,\label{x.11}\\
{\tilde \phi} &=& a +  \frac{1}{2} \ln \big(\frac{e^W}{e^W-1}\big),\label{x.12}
\end{eqnarray}
where $e^W >1$, i.e., $W >0$. We now define the transformation $e^W = {1}/{(1-\hat{r})}$ so that it requires that $0 <\hat{r}< 1$.
Applying this transformation and also using the linear transformation \eqref{x.5}, one can show that the solutions
\eqref{x.10}-\eqref{x.12} turn into the same form of the solution given by \eqref{x.6}-\eqref{x.8}, that is,
the solutions corresponding to both the valid ranges ${\tilde x}_2 +\frac{1}{2} e^{-2y} <0$
and ${\tilde x}_2 - \frac{1}{2} e^{-2y}>0$ can be expressed as the solution  given by \eqref{x.6}-\eqref{x.8} only with
$0 <\hat{r}< \infty$.

\bigskip

One can simply check that the solution  \eqref{x.6}-\eqref{x.8} does satisfy the equations \eqref{b.5}
with ${\tilde \Lambda} = -1/{l^2}$.  By considering this solution for the whole space-time, one sees that
the metric components \eqref{x.6} are ill behaved at $\hat{r} =0$  and $\hat{r} = 1$.
By looking at the scalar curvature, which is
${\tilde R} ~=~ {2(4 \hat{r} -7)}/l^2{\hat{r}^2}$,
we find that $\hat{r} =0$ is a curvature singularity. Note that the singularity at  $\hat{r} =0$ corresponds to the same true
singularity at the region ${\tilde x}_2 = \frac{1}{2} e^{-2y}$ which mentioned above.
We will see that $\hat{r} = 1$ is also an event horizon. The cross term  appeared in the metric is constant and thus for large $\hat{r}$ the metric is asymptotically
flat. However, the solution represents a stationary black string. We  wish to express the constant parameter $a$ and the radius $l$
in terms of the physical mass and axion charge per unit length of the black string. To calculate
the mass  we first need to have the asymptotic behavior of the solution given by \eqref{x.6}-\eqref{x.8}.
To this end, we set
\begin{eqnarray}
\hat{r} ~ = ~ \frac{2 l}{3} e^{2a} r, \label{x.13}
\end{eqnarray}
in equations \eqref{x.6}, \eqref{x.7} and \eqref{x.8} and only drop the hat sign from on the $\hat{t}$ and $\hat{x}$ coordinates.
Notice that it is not possible to similarly to the $t$ and $x$ coordinates  fix the
overall scaling of the $\hat{r}$  coordinate as $\hat{r}$ goes to infinity, since the metric asymptotically approaches
${l^2 d \hat{r}^2}/{4 \hat{r}^2}$. Therefore, for large $r$ the black string solution \eqref{x.6}-\eqref{x.8} approaches the  following
asymptotic solution
\begin{eqnarray}
{d {\tilde s}_{\eta}}^2 &=& -  d t^2 +   d {x}^2 + d \rho^2,\nonumber\\
{\tilde \phi} &=& -\frac{\rho}{l} +\frac{1}{2} \ln(\frac{3}{2 l}),~~~~~~~{{\tilde H}} = 0, \label{x.14}
\end{eqnarray}
where we have set $r=e^{2\rho /l}$. The solution \eqref{x.14} is a flat solution for equations \eqref{b.5}.
As it can be seen from \eqref{x.14}, the radius $l$ is related to the derivative of the asymptotic value of the
dilaton field.
Now, in order to calculate the mass per unit length  of the string we use the Arnowitt-Deser-Misner
(ADM) procedure which was been applied for obtaining the mass and charge of three-dimensional black strings
\cite{Horowitz1}, \cite{Steif1} (see, also, \cite{Witten2}).
In this way, to calculate the mass one needs to perturb  the metric
as ${G}_{_{\Upsilon\Lambda}} = {\eta}_{_{\Upsilon\Lambda}} + {\gamma}_{_{\Upsilon\Lambda}}$. Then, by integrating the
time-time component of the linearized form of the metric and dilaton contributions in equations  \eqref{b.5} over a spacelike surface,
the total mass is derived to be of the form\footnote{Note that the exponential contribution of dilaton field in action
\eqref{b.4} has been appeared as $e^{-2 \phi}$. Therefore, the formula  of the total mass \eqref{x.15} differs from those in Ref. \cite{Horowitz1}.} \cite{Horowitz1}, \cite{Steif1}:
\begin{eqnarray}
{\cal M}_{tot} ~ =~ \frac{1}{2} \oint e^{-2 { \phi} }
\Big(\partial^j {\gamma}_{ij} - \partial_i \gamma -2 \gamma_{ij} \partial^j { \phi}\Big) dS^i, \label{x.15}
\end{eqnarray}
where $i$, $j$ run over spatial indices and $\gamma$ is the trace of the spatial components of
${\gamma}_{_{\Upsilon\Lambda}}$, i.e., ${\gamma} = {\gamma}_{i}^{~i}$.
Finally, in three dimensions, the axion charge per unit length associated with the field strength $H$ is given by \cite{Horowitz1}, \cite{Steif1}
\begin{eqnarray}
Q ~ =~ \frac{1}{2} \oint e^{-2 { \phi} } \ast H, \label{x.16}
\end{eqnarray}
where $\ast$ denotes the Hodge dual.
Thus, by considering the metric ${\eta}_{_{\Upsilon\Lambda}}$ corresponding to the line element of equation \eqref{x.14} and by using the specific
form of ${\gamma}_{_{\Upsilon\Lambda}}$ for the solution \eqref{x.6} we carry out the above procedure. For the black string solution
\eqref{x.6}-\eqref{x.8}, the mass and axion charge  per unit length measuring at the $\rho = \infty$ end are therefore
\begin{eqnarray}
M &=& \frac{3}{2 l}  e^{-2 a}, \label{x.17}\\
Q &=& \frac{2}{\sqrt{3}~ l}  e^{-2 a}. \label{x.18}
\end{eqnarray}
Clearly, $M = \frac{3 \sqrt{3}}{4} Q$ and thus $M> Q$. Using these results, the final expression
for the black string solution is obtained to be
\begin{eqnarray}
{d {\tilde s}}^2 &=& - (1-\frac{2 M}{{r}}) d {t}^2 +  (1-\frac{2 M}{3 {r}}) d {x}^2 +
\frac{2}{\sqrt{3}}~ d {t} d {x} +(1-\frac{M}{{r}})^{-2} ~ \frac{l^2 d {r}^2}{4 {r}^2},\label{x.19}\\
{{\tilde H}_{{r} {t} {x}}} &=& \frac{3 Q}{2 {r}^2},\label{x.20}\\
{\tilde \phi} &=& -\frac{1}{2}\ln r +\frac{1}{2}  \ln(\frac{3}{2 l}). \label{x.21}
\end{eqnarray}
The metric \eqref{x.19} possesses  a single horizon at $r=M$ and a curvature singularity at $r=0$, since the scalar curvature is
${\tilde R} = [2M(4r-7M)]/l^2 r^2$. The  metric \eqref{x.19} appears to be  somewhat analogous to
the extremal limit $|{\mathbb Q}| = {\mathbb M}$ of the charged black string \eqref{b.20} \cite{Horowitz1}.
However there is one important difference which that is the appearance of the cross term in the metric.
The metric \eqref{x.19} also possesses two independent Killing vectors
$\partial / \partial t$ and $\partial / \partial t + \beta \partial / \partial x$ for $\beta \neq 0$.

$\bullet$~ The Killing vector $\partial / \partial t$ with the norm ${\tilde G}_{tt} = -1 + 2M/r$ becomes null at $r=2M$, which lies
outside the event horizon $r=M$. It is a time translational at infinity and becomes timelike for $r>2M$, while becomes
spacelike for $M \leq r < 2M$ and also inside the event horizon.

$\bullet$~ The Killing vector $\partial / \partial t + \beta \partial / \partial x$ becomes timelike for the
ranges $r< [2M(\beta - \sqrt{3})]/[3(\beta-1/\sqrt{3})]$ when $\beta \in (-\infty , -\sqrt{3})\cup (\sqrt{3} , \infty)$
which lies inside the event horizon, and
$r>[2M(\beta - \sqrt{3})]/[3(\beta-1/\sqrt{3})]$ when $\beta \in (-\sqrt{3} , 1/\sqrt{3})-\{0\}$. It is also timelike at infinity
when $\beta \in (-\sqrt{3} , 1/\sqrt{3})-\{0\}$. For  $\beta = -\sqrt{3}$ it becomes null, and for $\beta = \sqrt{3}$ it stays everywhere spacelike.

As mentioned above, our black string solution is asymptotically flat. Thus, the non-Abelian T-duality transformation changes
the asymptotic behavior from $AdS_3$ to flat. One question that is in our mind is that although the black string
metric \eqref{x.19} is  asymptotically flat, its near horizon geometry is of the  $AdS_3$ type, that is, in the limit
$r \rightarrow M$, one gets ${\tilde R} \rightarrow -6/l^2$ and ${\tilde R}_{_{\Upsilon\Lambda}} \rightarrow -2/l^2 {\tilde G}_{_{\Upsilon\Lambda}}$.
The study of near horizon behavior is important, because according to Strominger's proposal \cite{Strominger} one
asserts  that the statistical entropy of any black hole whose
near horizon geometry contains an $AdS_3$ factor can be calculated by using the statistical counting of
microstates of the BTZ black hole.
Lastly, we have to note the fact that the charged black string solution
\eqref{x.19}-\eqref{x.21} with $l=1$ is a dual solution for the $SL(2,\mathbb{R})$ WZW model.


\section{\large Discussion and Conclusion}

We have reviewed aspects of Poisson-Lie T-duality in the presence of spectator fields.
We have reobtained Buscher's duality transformations from the Poisson-Lie T-duality transformations in the presence of spectators.
Then using this approach we have studied
the Abelian T-dualization of the BTZ black hole solutions given by \eqref{b.2} in such a way that
our solutions  are in agreement with those of Ref. \cite{Horowitz2}.
We have  explicitly constructed a dual pair of  sigma models related by Poisson-Lie symmetry so that the original model has built
on a $2+1$-dimensional manifold ${\cal M} \approx O \times {\bf G}$ in which ${\bf G}$ is a  two-dimensional real non-Abelian Lie group
that acts freely on ${\cal M}$ and $O$ is the orbit of ${\bf G}$ in ${\cal M}$.
The metric of the model depends on a non-zero real constant parameter $k$ so that it
describes an anti-de Sitter space with $k>0$ while for $k<0$ we have a de Sitter space.
Then we have shown that the original model indeed is canonically equivalent to
the $SL(2,\mathbb{R})$ WZW model for a given value of the background parameters.
Therefore, we have shown that the Poisson-Lie T-duality relates the
$SL(2 , \mathbb{R})$ WZW model to a sigma model defined on a $2+1$-dimensional manifold ${\cal M}$.
In addition, by a convenient coordinate transformation we have shown that the original model
describes a string propagating in a spacetime with the BTZ black hole
vacuum metric.  In this way we have found a new family of the solutions to low energy string theory with the BTZ black hole
vacuum metric, constant dilaton field and a new torsion potential.
The dual model has built on a $2+1$-dimensional target
manifold $\tilde {\cal M}$ with two-dimensional real Abelian Lie group ${\tilde {\bf G}}$ acting freely on it.
We have shown that the dual model yields a new three-dimensional charged black string which is stationary and asymptotically flat.
In this way,  the non-Abelian T-duality
transformation  has  related a solution with no horizon
and no curvature singularity  to a solution with a single horizon and a curvature singularity.
In addition to,  it  changes
the asymptotic behavior of solutions from $AdS_3$ to flat.
According to our findings, it seems that under the non-Abelian T-duality transformation the mass and charge
have not restored from the dual model to the original one.
We have thus found a new family of solutions in the abstract is rather too strong.
We have also  investigated the effect of Poisson-Lie T-duality on the singularities of spaces of T-dual models and have shown
that this duality takes those singular regions to regular regions as was the case with the 2D black holes
\cite{Dijgkraaf} and 3D black strings \cite{Horowitz1},  \cite{Steif1}, \cite{Steif2}.

Nevertheless, in order to address the question of the non-Abelian T-dualization of the BTZ black hole solutions one has to show
that the BTZ black hole metric has sufficient number of independent Killing vectors. Then the isometry subgroup of
the metric can be taken as one of the subgroups of the Drinfeld double. In order to satisfy the dualizability conditions the other
subgroup must be chosen Abelian. The isometry groups of metrics can be used for construction
of non-Abelian T-dual backgrounds. In the present case, to construct the  dualizable metrics by the Poisson-Lie T-duality one needs
a three-dimensional subalgebra of Killing vectors that generates group of isometries which acts freely and transitively on the
three-dimensional manifold ${\cal M}$ where the BTZ metric is defined.
Thus one can construct several non-Abelian T-dual backgrounds for the BTZ  metric.
We intend to address this problem in the future.

\bigskip

\noindent {\bf Acknowledgments:}
We  would like thank the referee for comments that helped us improve this issue.
A. Eghbali is especially grateful to S. Hoseinzadeh for sharing some of his insights with
him. This work has been supported by the research vice chancellor of Azarbaijan Shahid Madani University of Iran under research fund No. 95.537.


\end{document}